# A generic attack to ciphers

Li An-Ping

Beijing 100080
apli0001@sina.com

Abstract

In this paper, we present a generic attack for ciphers, which
is in essence a collision attack on the secret keys of ciphers .

In cryptography there were more researches formerly on the collision attacks mainly for hash functions in preventing falsification, see [1], [2] and [3]. Actually, the collision attacks also exist for ciphers. In this paper, we present a generic attack to ciphers, which is essentially a collision attack on the secret keys of ciphers.

Suppose that $Crypto(key, plaintext)$ is a cipher with $|key|= n$. Let $x_0$ be a plaintext of $n$ bits, then the range of space $Crypto(key, x_0)$ ( taken as the function of variable $key$ ) will be about as large as the one of the key space if function $Crypto(key, x_0)$ behaves as random one.

Let $X_0$ be a $3n$-bits string of plaintext, we will show later that with very high probability, $Crypto(key_1, X_0) \neq Crypto(key_2, X_0)$ for two distinct keys $key_1$ and $key_2$.

Arbitrarily take $N$ distinct keys $key_i, 1 \leq i \leq N$, and get ciphertexts $Crypto(key_i, X_0)$, $1 \leq i \leq N$, to make a dictionary $\mathcal{D}$ with the entrice $(Crypto(key_i, X_0), key_i)$, $1 \leq i \leq N$, arranged in the order of the first component as a number. In the case there are a set of keys, $key_{i_1}, key_{i_2}, \ldots, key_{i_s}$ such that they have same imagine

$$Y' = Crypto(key_{i_1}, X_0) = Crypto(key_{i_2}, X_0) = \ldots = Crypto(key_{i_s}, X_0),$$

then modify the entry as $(Y', (key_{i_1}, \ldots, key_{i_s}))$.

**Proposition 1** It is assumed that it is equal-probability for the function $Crypto(key, X_0)$ to take each ciphertext, denoted by $p$ the probability that the ciphertexts $Crypto(key_i, X_0)$, $1 \leq i \leq N$, are different each other, then

$$p > 1 - \frac{1}{2^n}.$$

Proof. Clearly,

$$p = \frac{\prod_{0 \leq k < N}(2^{3n} - k)}{2^{3nN}} \geq \prod_{1 \leq k < N} \exp(-(\frac{k}{2^{3n}} + \frac{k^2}{2^{6n}}))$$

$$\geq \exp(-(\frac{(N-1)N}{2 \cdot 2^{3n}} + \frac{N^3}{3 \cdot 2^{6n}}))$$

$$> 1 - \frac{(N-1)N}{2 \cdot 2^{3n}} - \frac{N^3}{3 \cdot 2^{6n}}$$

$$> 1 - \frac{1}{2^n}.$$

Suppose an adversary get a ciphertext $Crypto(key^*, X_0)$ for some key $key^*$ to be found, and find $Crypto(key^*, X_0) = Crypto(key_i, X_0)$ for some $i, 1 \leq i \leq N$, then he will infer that $key^* = key_i$ with very high successful probability by Proposition 1. Let $N = 2^m$, we know it will need at most $3nm$ comparisons to determine/find $Crypto(key^*, X_0)$ in the dictionary $\mathcal{D}$.

Moreover, suppose that the adversary get $2^t$ ciphertexts $Crypto(key^{(i)}, X_0)$, $1 \leq i \leq 2^t$, denoted by $\hat{p}$ the probability that there is at least one of ciphertexts $Crypto(key^{(i)}, X_0)$, $1 \leq i \leq 2^t$, in the dictionary $\mathcal{D}$. It has that

**Proposition 2** If $t \geq n - m + 2$, then $\hat{p} \geq 0.98$.

Proof. It is obvious that

$$\hat{p} \geq 1 - \left(\frac{2^n - 2^m}{2^n}\right)^{2^t} \geq 1 - 1/e^{2^{t+m-n}}.$$

The conclusion is followed directly from the formula above.

The result above is somewhat similar to the birthday paradox.

In the capability of the modern computers, it seems that the size of the dictionary $\mathcal{D}$ now has been able to achieve about 50 bits, so the analysis above suggests that it should use a larger length of secret keys for a secure cryptosystem.

The analysis above is a generic one, which may be applied to block ciphers, stream ciphers and MAC. Besides, it needs only a few of ciphertexts for one key, e.g. $n = 128$, then $X_0$ is a 48-bytes of string. This makes the attack more feasible.

It maybe should be mentioned that for a good cipher $Crypto(key, plaintext)$, the number of pre-imagines should not be too much, for otherwise the plaintext would be easy recovered.

# References

[1] A. Menezes, P. van Oorschot, S. Vanstone, Handbook of Applied Cryptograpgy, CRC Press, 1997.

[2] B. Preneel, Analysis and design of cryptographic hash functions, PhD thesis, Katholieke Universiteit Leuven (Belgium), Jan. 1993.

[3] X. Wang, H. Yu: How to break MD5 and Other Hash Functions, Advances in EUROCRYPT 2005, LNCS 3494, pp. 19-35, 2005.